\documentstyle[twocolumn,aps,prl,epsf]{revtex}

\def\tr{{\rm\,tr\,}}

\newcommand{\ket}[1]{|#1\rangle}

\newcommand{\half}{\textstyle{\frac{1}{2}}}
\begin{document}
\draft
\title{ Dimer Decimation and Intricately Nested Localized-Ballistic Phases of Kicked Harper}
\author{Toma\v z Prosen$^a$, Indubala I. Satija$^b$ and Nausheen Shah$^b$}
\address{
(a) Physics Department, Faculty of Mathematics and Physics,
Univ.of Ljubljana, Jadranska 19, SI-1111 Ljubljana, Slovenia\\
(b) Department of Physics, George Mason University, Fairfax, VA 22030}
\date{\today}
\maketitle
\begin{abstract}
Dimer decimation scheme is introduced in order to study the kicked 
quantum systems exhibiting localization transition. The tight-binding
representation of the model is mapped to a vectorized dimer 
where an asymptotic dissociation of the dimer is shown to correspond 
to the vanishing of the transmission coefficient thru the system. 
The method unveils an intricate nesting of
extended and localized phases in two-dimensional parameter space. 
In addition to computing transport characteristics with extremely high precision,
the renormalization tools also provide a new method to compute quasienergy spectrum.

\end{abstract}
\pacs{PACS numbers: 72.15.Rn+72.15-v}

\narrowtext

Kicked Harper model has emerged as an important model in the
recent literature on quantum chaos\cite{Lima,rev,Roland}. 
The model exhibits both ballistic/extended as well as localized states 
in the regime where the corresponding classical system is chaotic. 
This challenges the concept of dynamical localization in non-integrable 
systems (such as kicked rotor), interpreted as the suppression of 
quantum diffusion due to classical chaos. 
Localization-delocalization transition of the model although has its 
roots in the corresponding un-kicked case is very complex with mixed 
spectrum and nontrivial multi-fractal characteristics\cite{rev,Roland}.
We use renormalization group (RG) techniques to study this richness and 
complexity.  
Our methodology has clear conceptual and numerical 
advantages over earlier methods for spotting critical states and locating
localization transitions. 

The kicked Harper model\cite{Lima} is described by the time-dependent Hamiltonian
\begin{equation}
H(t)=L\cos(p)+K \cos(q) \sum_{k=-\infty}^\infty \delta(t-k) .
\label{KHamil}
\end{equation}
Here $q,p$ is a canonically conjugate pair of variables on a cylinder.
In the absence of kicking, the quantum system $H_0=L\cos(p)+K \cos(q)$, 
can be written as a nearest-neighbor (nn) lattice model.
We take $p = m\hbar$, and $e^{iq}$ as a translational operator for $p$,
obtaining the well known Harper equation,\cite{Harper}
\begin{equation}
\half K(\psi_{m+1}+ \psi_{m-1} ) + L \cos(\hbar m) \psi_m = \epsilon \psi_m
\label{HE}
\end{equation}
If we choose $\hbar=2\pi\sigma$, where $\sigma$ is a Roth
number, the model exhibits localization-delocalization transition
at $K=L$.\cite{Harper} This transition has been analyzed by various 
RG schemes.\cite{Ostlund,KSrg,GS}

For the Hamiltonian (\ref{KHamil}) that is periodic in time, 
the matrix elements of the evolution operator ${\bf U}$, in the  
angular-momentum basis $\ket{m}$ with eigenvalues $p=m\hbar$, read
\begin{equation}
U_{m,m\prime}=\exp[-2i\bar{L}\cos(m\hbar)] (-i)^{m-m\prime} 
J_{m-m\prime}(2\bar{K})
\label{propagator}
\end{equation}
Here $J_r$ is a Bessel function of
order $r$ and $\bar{K}=K/(2\hbar)$ and  $\bar{L}=L/(2\hbar)$.
The spectral problem 
${\bf U}\ket{\psi_\omega}=e^{-i\omega t}\ket{\psi_\omega}$, ($\omega$ being the quasienergy)
involving unitary matrix of the above form was transformed\cite{QC}
to a {\em real} lattice tight binding 
model (tbm) with angular momentum quantum number representing 
the lattice index
\begin{equation}
\sum_{r=-\infty}^{\infty} B^m_r u_{m+r}= 0,
\label{Beq}
\end{equation}
where the coefficients $B^m_r$ are  
\begin{equation}
B^m_r = J_r(\bar{K}) \sin[\bar{L}\cos(m\hbar)-\pi r/2 -\omega/2].
\label{Bs}
\end{equation}
For finite $\bar{K}$, the tbm (\ref{Beq}) effectively contributes only 
few terms as Bessel's function exhibit fast decay when $|r| > |\bar{K}|$.
Therefore, the tbm describes a lattice model with a finite range of 
interaction denoted as $b$ ($b\approx\bar{K}$). In the limit of small 
$\bar{K},\bar{L},\omega$, tbm reduces to the simple Harper equation 
(\ref{HE}) with $\epsilon=\hbar\omega$\cite{Harper}. 

The renormalization tools have proven to be a great asset in studying 
localization transitions in aperiodic one-dimensional (1D) nn lattice 
models. We use RG methodology to investigate localization transitions 
in the kicked system (\ref{KHamil}) represented by the lattice model 
(\ref{Beq},\ref{Bs}).  
We propose two independent RG schemes, which are generalization of {\it dimer decimation} scheme
of the nn case\cite{GS} which reduce the lattice 
problem to a (vector, or block) dimer, the discrete analog of the 
textbook example of the quantum barrier problem. The transport characteristics of the lattice model
can be understood as due to quantum interference within the dimer.
Furthermore, the decay of the coupling 
between the two sites of the dimer is shown to correspond to the 
vanishing of the transmission probability: i.e, the localized phase is 
asymptotically a broken dimer under the RG flow.
As we discuss below, scaling analysis of the transmission properties 
provide an extremely accurate tool to distinguish extended,
critical and localized phases. 
The key idea underlying both renormalization schemes is the 
{\it simultaneous} decimation of the two central sites of the doubly 
infinite lattice $-\infty, ... -2,-1,0,1,2,....\infty$, namely $\pm 1$,
$\pm 2$ and so on, after we have eliminated the site $m=0$.
Both can be applied to general tbm (\ref{Beq}) with arbitrary 
coefficients $B^r_m$.
 
In the first scheme, which we will refer to as {\em vector decimation},
the decimation is done on a vectorized form of the tbm:
the lattice model (\ref{Beq}) is first transformed to a 
{\em nn vector model} where each site is associated
with a $b$-dimensional vector $\Phi_m$: 
$\Phi_m = ( X^0_m, X^1_m, ....X^{b-1}_m )$ where
$X^0_m = u_m$ and for $0 < r < b$,
\begin{equation}
X^{r}_m =B^m_{b-r} u_m+\sum_{k=0}^{r-1} B^m_{b-k} [u_{m+r-k}+u_{m-r+k}] 
\end{equation}
Here we assumed, for simplicity, that the lattice model
exhibits reflection symmetry $B^m_r\equiv B^m_{-r}$.
The lattice model (\ref{Beq}) can now be written as a nn vector tbm,
\begin{equation}
\Phi_{m-1}+\Phi_{m+1}- {\bf V}_m\Phi_m=0,
\label{VTBM}
\end{equation}
where ${\bf V}_m$ is a $b \times b$ matrix whose
non-zero elements are given by,
$V_m(k,k+1) = 1$,
$V_m(k,k)= -B^m_{b-1}\delta_{k,b}$,
$V_m(k,b) = B^m_{k+1}-B^m_{k-1}$
and $V_m(b-1,b) = 2B^m_b-B^m_{b-2}$. 
These elements have been normalized with $B^m_r$.
%%% TP: I do not understand the meaning of the above statement???
The dimer decimation on this vector tbm is described by the 
vectorization of the RG flow of the nn tbm\cite{GS}. 
We first eliminate the central site ($m=0$) and then carry out the iterative process of 
decimating the two central sites.
At the $n^{\rm th}$ step where all sites with $|m| < n$ have been
eliminated, the tbm for $m=\pm n$ can
be written as
\begin{eqnarray}
\Phi_{n+1} + {\bf G}(n) \Phi_{-n} - {\bf E}(n) \Phi_{n}&=&0\\
\Phi_{-n-1} +{\bf G}(n) \Phi_{n} - {\bf E}(n) \Phi_{-n}&=&0  \label{RGv}
\end{eqnarray}
where ${\bf G}(n)$ and ${\bf E}(n)$ are $b \times b$ matrices.
With initial conditions ${\bf G}(1)={\bf V}_0^{-1}$ and
${\bf E}(1) = {\bf V}_1 - {\bf G}(1)$, obtained by decimating the central site, 
the renormalized matrices ${\bf G}(n)$ and ${\bf E}(n)$
are given by the following RG flow, the matrix version of 
the dimer map\cite{GS}
\begin{eqnarray}
{\bf G}(n+1)&=& [{\bf E}(n) {\bf G}^{-1}(n) {\bf E}(n) - 
{\bf G}(n)]^{-1}\\
{\bf E}(n+1) &=& V_{n+1} + [{\bf G}(n) {\bf E}^{-1}(n) {\bf G}(n) 
- {\bf E}(n)]^{-1}\nonumber
\end{eqnarray}

In the second scheme, which we will refer to as {\em scalar decimation}, 
we seek a renormalization scheme which {\em preserves the banded form 
of the tbm} (\ref{Beq}). 
%It turns out that decimation of site $m=0$ or any further simultaneous 
%decimation of the two central sites of the doubly infinite lattice 
%can again be described by a banded matrix of bandwidth $2b+1$ having 
%additional $2b\times 2b$ block matrix nested in the center.
We define the $2b\times 2b$ {\em central matrix} ${\bf A}$ with an 
initial value $A_{j,k}(0) := \bar{B}^j_{k-j}$,
$j,k\in\{-b,\ldots,-2,-1,1,2,\ldots,b\}$,
where $\bar{B}^m_r := B^m_r - B^m_{-m}B^0_{m+r}/B^0_0$
describes the renormalized tbm after only the central site $0$ has been 
decimated from eqn.(\ref{Beq}).
Note that expressing $u_{\pm 1}$ from the central two eqs. and substituting 
for $u_{\pm 1}$ in the remaining eqs. one obtains tbm on a decimated lattice 
$\pm 2,\pm 3\ldots$ of the same band structure with a renormalized 
$2b\times 2b$ matrix $A_{j,k}(1)$ in the center. The recursive scheme 
describing the $n^{\rm th}$ step renormalization of the matrix ${\bf A}$
is
\begin{eqnarray}
A_{j,k}(n+1) &=& \det{\bf L}_{j,k}(n)/D(n), \label{RGs} \\
{\bf L}_{j,k}(n) &=& 
\pmatrix{A'_{j^+,k^+}(n) & A'_{j^+,-1}(n) & A'_{j^+,1}(n) \cr
         A'_{-1,k^+}(n)  & A_{-1,-1}(n)   & A_{-1,1}(n) \cr
         A'_{1,k^+}(n)   & A_{1,-1}(n)    & A_{1,1}(n) \cr} \nonumber
\end{eqnarray}
where $D(n):=A_{-1,-1}(n)A_{1,1}(n) - A_{-1,1}(n)A_{1,-1}(n)$ is a central 
$2\times 2$
determinant, $j^+$ denotes the renormalized lattice site: 
$j^+:=j+1,j-1$ if $j > 0,< 0$, respectively, and
$$
A'_{j,k}(n):= 
\cases{
A_{j,k}(n); &  $j,k\in\{-b,\ldots,-1,1,\ldots,b\}$  \cr
B^{j+n}_{k-j}; & $j > b$ or $k > b$ \cr
B^{j-n}_{k-j}; & $j <-b$ or $k < -b$ 
}
$$
Comparing vector and scalar decimation flows, the later although may lack 
the elegance formulation of the vector decimation, has various numerical advantages in addition to being very fast,  
as the only source of possible singularity, namely the central  $2\times 2$ 
determinant $D(n)$, is easy to control, and the whole procedure is 
completely stable against oversizing the bandwidth $b$. However, two independent methods
provides a unique advantage in confirming  many subtle features of the phase diagram that are discussed below.

The important quantity that characterizes the transport properties
is the effective coupling of the renormalized dimer. 
It is the ratio $R$ of the off-diagonal to the diagonal
part of the renormalized lattice. In the vector and scalar decimation, 
the appropriate quantities are
$R_v = \tr({\bf G} {\bf G}^{\dag})/\tr({\bf E} {\bf E}^{\dag})$
and $R_s = \tr({\bf A_-}{\bf A_-}^{\dag})/\tr({\bf A_+} {\bf A_+}^{\dag}) $
where we name the offdiagonal and diagonal $b\times b$ blocks 
respectively as $(A_-)_{j,k} := A_{j,-k}$, and 
$(A_+)_{j,k}=A_{j,k}$, $j,k\in\{1,\ldots,b\}$ 
(assuming reflection symmetry $A_{n,l}\equiv A_{-n,-l}$).

To confirm that the parameter $R$ is indeed related to the transmission 
coefficient of the model, we have done direct calculation of the 
transmission properties by solving the scattering problem on a momentum 
lattice for a truncated kicked model. 
This is achieved by replacing kinetic energy term by 
$L\cos(\hbar m)\theta(M-|m|)$, $\theta(n\ge 0) := 1,\theta(n < 0) := 0$, 
where the parameter $M$ defines the size of the {\em scattering region} 
$|m| \le M$. This scattering model is equivalent to tbm (\ref{Beq},\ref{Bs}) 
where $\bar{L}$ is replaced by $\bar{L}\theta(M-|m|)$.
Outside the scattering region, $m > |M|$, the wave-function is a 
superposition of properly normalized plane waves 
$\psi^{\pm(l)}_m = |\sin(\kappa_l)|^{-1/2}\exp(\pm i\kappa_l m)$, with
$\cos(\kappa_l) = (\omega + 2\pi l)/(2\bar{K})$, $l$ integer.
The reflection and transmission matrices ${\bf R}$ and ${\bf T}$ are 
determined by matching the ans\" atze for the asymptotic solutions,
$u_m = \psi^{+(l)}_m + \sum_{l'}R_{l l'}\psi^{-(l')}_m$, for $m <-M$,
and $u_m = \sum_{l'}T_{l l'}\psi^{+(l')}_m$, for $m >-M$, on tbm 
(\ref{Beq}) for $|m| \le M$. Importantly, decimation scheme makes the 
solution of the scattering problem for large $M$ very efficient, as $n=M-b$ 
iterates of the RG map (\ref{RGs}) are performed first in order to maximally 
reduce the size of the {\em scattering region}, as ${\bf A}(n)$ 
{\em does not depend on truncation} for $n \le M-b$. Our numerical 
calculations confirmed that results obtained using the total transmission 
probability $P(M=n+b) := \sum_{l,l'} |T_{l l'}|^2$ are fully consistent 
with the ones based on the ratios $R_v(n)$ and $R_s(n)$. 

It turns out that the scaling exponent 
$\beta = \lim_{n\to\infty}\beta(n)$, $\beta(n)=\log S(n)/\log n$ 
(where $S=P$, $R_s$, or $R_v$) provides a very effective means to 
describe transport properties. The 
extended states are 
described by (typically monotonic) convergence of $\beta(n)$ $\to 0$. In case of exponential localization,
$\beta(n) \to -\infty$.
The decay $S(n)\sim\exp(-2n/\xi)$ can be used to calculate the localization
length $\xi$. 
In contrast, the critical states are characterized by 
negative $\beta$ exhibiting non convergent, oscillatory behavior. 
Therefore, asymptotically, both the critical and the localized kicked model
describes a broken dimer.
In fig.1, we display RG flow at the 
Fibonacci iterates $F(f)$, 
the successive denominators of the continued fraction approximation of 
$\sigma=(\sqrt{5}-1)/2$, which is kept fixed throughout the paper. However,
the method can be implemented for arbitrary $\sigma$,
By iterating RG equations upto $F_{33}=5,702,887$, the localization 
thresholds are determined almost to machine precision. 
Most of our studies describing the
variation in the transport properties are carried out for $\omega=0$ ,
which appears to be the eigenvalue for all parameters.state. 
However, as we discuss later, the RG methodology can also be used
to compute the quasienergy spectrum and analyze its transport characteristics.

\begin{figure}
\begin{center}
\leavevmode
\epsfxsize=3in
\epsfbox{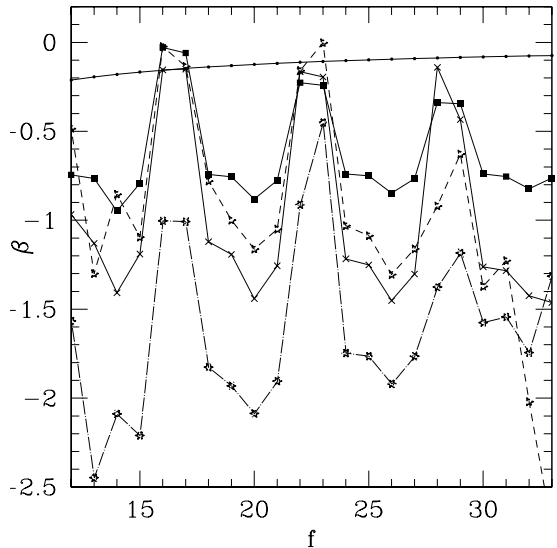}
\end{center}
\protect\caption{ The RG flow at Fibonacci iterates, $F(f)$.
The monotonic curve with dots describes extended states for
$\bar{K}=4$ and $\bar{L}=1$.
The oscillatory pattern characterizing the critical states correspond to
unkicked model (diamond) ,
kicked model with $\bar{K}=\bar{L}=4$ (crosses). In addition, critical state off the symmetry line is shown for
$\bar{K}=4.4$, $\bar{L}=2.516711$ and its dual
(short and long dashed lines with stars).}
\label{fig1}
\end{figure}

We confirm the critical nature of the state along the symmetry axis $K=L$ 
where the exponent $\beta$ is always found to be negative and bounded.
For any value of $K$, RG flow displays the oscillatory pattern shown in 
fig.1 with only variation being the amplitude of the oscillation. 
Fig.2 shows variation in $\beta$ with $K$ where smooth variations 
are intermitantly intrupted by rapid variations. As we discuss below, 
this may be related to the existence of critical manifolds off the symmetry line.

\begin{figure}
\begin{center}
\leavevmode
\epsfxsize=3in
\epsfbox{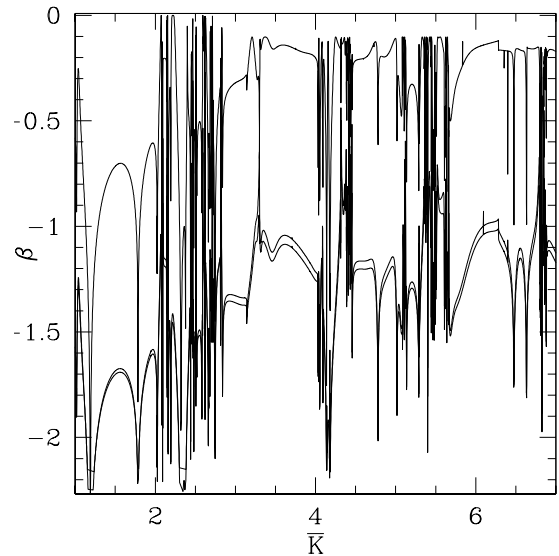}
\end{center}
\protect\caption{Variations in $\beta$ along the line $K=L$ in kicked Harper. 
Three RG iterates corresponding $F_{23}$, $F_{24}$
and $F_{25}$ (darker line) are shown.}
\label{fig2}
\end{figure}

One of the most remarkable and intriguing result of the RG analysis is the 
cascade of localization-delocalization transitions 
{\it off the symmetry line} where the extended and the localized regimes
display intricate pattern in two-dimensional parameter space as shown in 
fig.3. The extended phase of the unkicked model ($K > L$ regime)
is landscaped by various patches of localized regimes and similar
behavior is seen in the localized phase consisting of patches of extended 
regime which is the {\em dual} image of the extended reentries. 
A detailed numerical analysis suggests that the critical manifolds in 
$(K,L)$ space, corresponding to singular continuous states, can only correspond 
to the localization-delocalization boundary, and hence
we {\em do not } find any {\em fat critical phase}.
Furthermore, comparing figs. 2 and 3, the rapid variations in $\beta$ on $K=L$ 
may have its origin in collisions of the critical manifold with itself on the 
symmetry line.

Scanning one-parameter space, say by varying parameter $K$,
one sees a cascade of reentrant transitions where ballistic (localized)
transport reappears and survives in a finite
window in parameter after becoming localized (extended) as shown in fig.4.
We believe that these ''transitions´ are related to the cascades of
transitions predicted by semiclassical methods\cite{SS}.
The reentrancies exist for a band of quasiperienergy states and hence will
result in modulations in transport properties for a wave packet consisting of a
superposition of many quasienergy states.

\begin{figure}
\begin{center}
\leavevmode
\epsfxsize=3in
\epsfbox{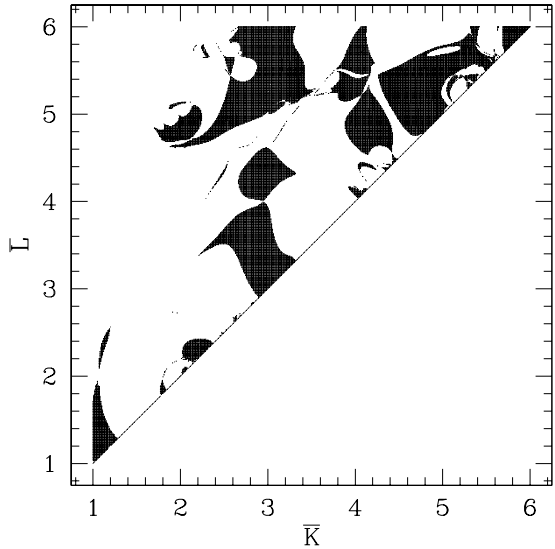}
\end{center}
\protect\caption{Two-dimensional phase diagram for $\omega=0$ state where the shaded parts describe extended (localized)
regimes for $L > K$ ( $K > L$). Almost all intricate details have a dual image and this provides an additional
confirmation of the numerics.}
\label{fig3}
\end{figure}

\begin{figure}
\begin{center}
\leavevmode
\epsfxsize=3in
\epsfbox{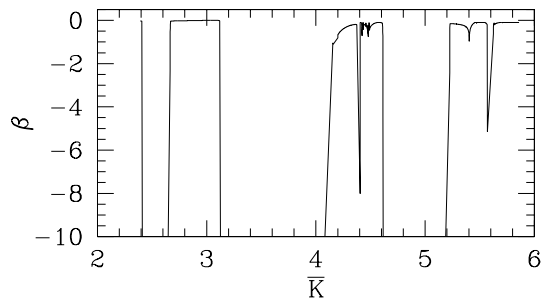}
\end{center}
\vspace{-1.4in}
\protect\caption{Series of localization-delocalization transitions indicated by $\beta$ changing from $-\infty$ to $0$
for a fixed $\bar{L}=4.2$ using $F_{25}$.
All transitions seen in this figure can be mapped with those shown in figure 3.}
\label{fig4}
\end{figure}

Treating $\omega$ as a running parameter , 
RG flow together with the {\em duality transformation} $(K,L) \rightarrow (L,K)$ 
(which maps extended states to 
localized states and vice versa as confirmed by detailed numerics) can also facilitate an accurate method 
to determine the quasienergy spectrum of the kicked Harper and related models. ( see fig.5) 
In RG scheme, it is in general difficult to distinguish 
localized and forbidden values of $\omega$ as it requires knowledge of $d\beta/d\omega$. 
However, by exploiting duality, localized spectrum can be easily separated from the 
spectral gaps as the later corresponds to 
$\beta(n)\to -\infty$ as well as $\beta^{\rm dual}(n)\to -\infty$. 
\begin{figure}
\begin{center}
\leavevmode
\epsfxsize=3in
\epsfbox{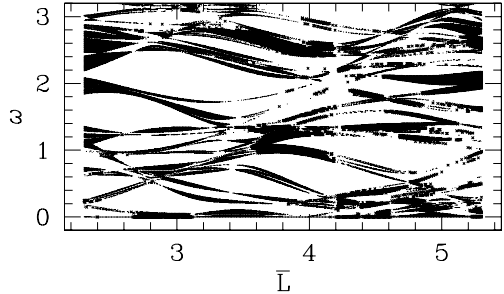}
\end{center}
\vspace{-1.4in}
\protect\caption{ Quasi-energy spectrum for
fixed $\bar{K}=4.2$. Lighter dots represent extended states. Darker crosses
are the localized states which were obtained as the extended states of the dual model.  
The localization-delocalization transitions for
various quasienergy states are clearly seen. For
$\omega=0$ state, the transitions can be mapped to those
of fig. 3 and the dual of the fig.4.}
\label{fig5}
\end{figure}

In summary, RG tools uncover many complex features of the
kicked model which would have been impossible with previously used methods. 
The  dimer decimation approach can be applied to a variety of problems which 
include unitary models without reflection symmetry
and can also be implemented to compute spectrally global quantities.
%such as infinite temperature spin stiffness of certain quantum spin chains
We hope that these tools will provide a new direction in resolving 
various important issues underlying the frontiers of localization 
phenomenon in complex systems.

The research of IIS is supported by National Science
Foundation Grant No. DMR~0072813. TP acknowledges Ministry of Science and 
Technology of Slovenia for a financial support.

\end{document}